\documentclass[12pt]{iopart}
\usepackage{graphicx}
\begin{document}
\title{A Stark decelerator on a chip}
\author{Samuel A. Meek, Horst Conrad, and Gerard Meijer}
\address{Fritz-Haber-Institut der Max-Planck-Gesellschaft, 
Faradayweg 4-6, 14195 Berlin, Germany}
\ead{meek@fhi-berlin.mpg.de}
\begin{abstract}
A microstructured array of 1254 electrodes on a substrate has been configured to generate an array of local minima of electric field strength with a periodicity of 120 $\mu$m about 25 $\mu$m above the substrate. 
By applying sinusoidally varying potentials to the electrodes, these minima can be made to move smoothly along the array. 
Polar molecules in low-field seeking quantum states can be trapped in these traveling potential wells. 
Recently, we experimentally demonstrated this by transporting metastable CO molecules at constant velocities above the substrate [Phys. Rev. Lett. 100 (2008) 153003]. 
Here, we outline and experimentally demonstrate how this microstructured array can be used to decelerate polar molecules directly from a molecular beam. 
For this, the sinusoidally varying potentials need to be switched on when the molecules arrive above the chip, their frequency needs to be chirped down in time, and they need to be switched off before the molecules leave the chip again. 
Deceleration of metastable CO molecules from an initial velocity of 360 m/s to a final velocity as low as 240 m/s is demonstrated in the 15-35 mK deep potential wells above the 5 cm long array of electrodes. 
This corresponds to a deceleration of almost $10^5$ $g$, and about 85 cm$^{-1}$ of kinetic energy is removed from the metastable CO molecules in this process.
\end{abstract}
\submitto{\NJP}
\maketitle

\section{Introduction}

The manipulation of ultracold atoms above a substrate using magnetic fields produced by microstructured current carrying wires on the substrate is a very active area of research right now. 
Present day micro-electronics technology enables multiple tools and devices to be integrated onto a compact substrate, thereby producing a wide variety of so-called atom chips \cite{folman02}.
Applications of atom chips are expected in many areas and these chips are, for instance, already used for matter-wave interferometry and precision-force sensing \cite{fortaugh07}.
When the microchip is made to form one wall of the vacuum chamber, a truly portable system can be obtained. 
Bose-Einstein condensation of $^{87}$Rb atoms on a microchip has been demonstrated in such a portable vacuum system, and it is anticipated that a chip-based BEC-compatible vacuum system can occupy a volume of less than 0.5 L. \cite{du04}.

Via miniaturization, strong electric field gradients can be produced as well. 
This has been used to push atoms along a magnetic guide on a chip \cite{kruger03} and to demonstrate electrodynamic trapping of spinless atoms in a microscopic volume on an atom chip \cite{kishimoto06}.
The strong electric field gradients and the possibility to engineer complicated electric field structures on a chip are also ideal for the precise manipulation of polar molecules. 
The latter is required, for instance, to implement proposed schemes of quantum computation that use polar molecules as qubits \cite{demille02,andre06}.

There are two major obstacles that make experiments with molecules on a chip less straightforward than with atoms. 
First, the methods to load molecules onto the chip are less matured than for atoms. 
It is more difficult to produce dense samples of quantum-state selected cold molecules than it is for atoms, and these samples then still need to be transported onto the chip. 
Several experimental schemes that have been demonstrated to produce samples of trapped, cold molecules start with a molecular beam. 
The molecules in the beam are decelerated by the use of, for instance, electric, magnetic or optical fields, and are subsequently trapped \cite{meerakker08}. 
Rather than transporting these stationary, cold samples to the chip, we have demonstrated that one can load the molecules onto a chip directly from a molecular beam, provided the traps on the chip move along with the molecules in the beam \cite{meek08}.
Recently, spectacular progress has been made in the production of samples of ultracold heteronuclear alkali dimers in their ro-vibrational ground state level \cite{deiglmayr08,ni08}.
It is anticipated that these samples of polar molecules can be loaded from the optical trap in which they are held now into electric field traps above a chip in the not too distant future. 
Moreover, polar molecules might also be loaded onto a chip by implementing such a chip in a cryogenic buffer gas cell \cite{weinstein98}, as the relatively low voltages that are required for trapping on a chip might be compatible with the conditions in this cell.
A second obstacle for the experiments with molecules on a chip is that detection methods for molecules are (much) less sensitive than those for atoms. 
As molecules in general lack a closed two-level system, efficient detection using absorption or laser induced fluorescence is not possible; whereas a single alkali atom can scatter up to 10$^7$ photons per second, a molecule normally ends up in a different quantum-state -- and turns dark -- after scattering only a few photons. 
In vacuum, molecules can normally be most sensitively detected using ionization based detection schemes, but these schemes are difficult to implement close to the microchip. 
Detection of molecules on a chip using micro-cavities, like is done for atoms \cite{trupke07}, might be an option, although even there, molecules have the disadvantage that a single ro-vibrational transition in an electronic band is generally weaker than an electronic transition for atoms, and for many molecules, the electronic transitions are in the more difficult to access near-UV region of the spectrum. 
In the experiments we have performed so far, we have circumvented this second obstacle by letting the molecules come off the chip again and by using sensitive detection schemes that have been developed for molecules in free flight in a molecular beam \cite{meek08}.

In this paper, we present the loading of polar molecules from a supersonic beam into traveling potential wells on a 5 cm long chip. 
These traveling potential wells originally move with the mean velocity of the molecules in the beam, but their velocity is then chirped down to two-thirds of the original velocity while the molecules are on the chip. 
Metastable CO molecules with an initial velocity of 360 m/s are captured in the approximately 15-35 mK deep potential wells above the chip and, after more than half of their kinetic energy has been removed, leave the chip again to be detected further downstream. 

In the following, the operation principle of the Stark decelerator on the chip is discussed, and a detailed description of the experimental setup is given. 
The experimental results are presented and compared to the outcome of trajectory calculations, before the paper concludes with a discussion of the future prospects of this miniaturized new device.  

\section{Operation principle}

\begin{figure}
\center{\includegraphics[scale=0.8]{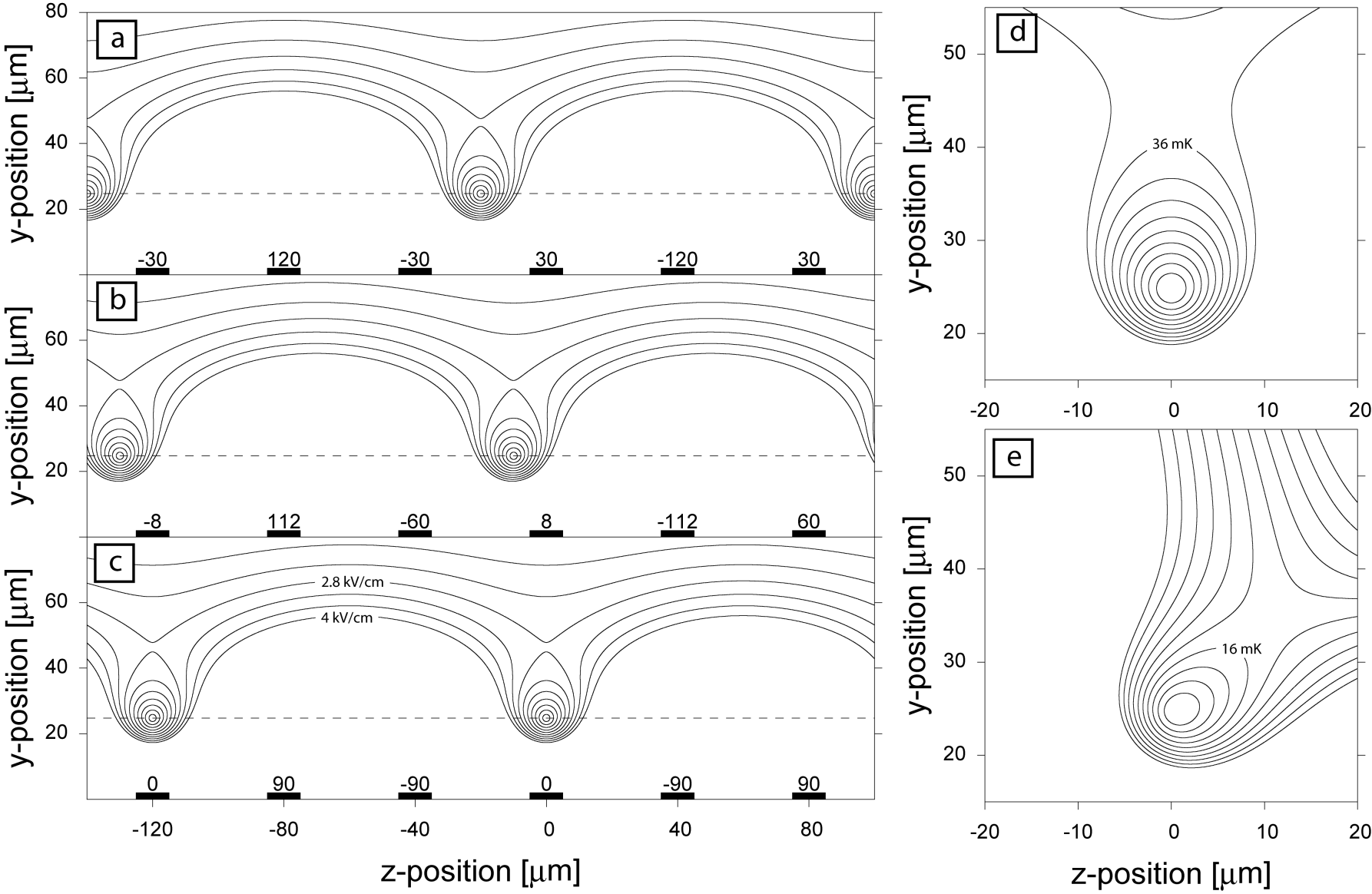}}
\caption{
(a) - (c): Calculated contour lines of equal electric field strength above a periodic array of electrodes, displayed at an interval of 0.4 kV/cm. 
The position of the electrodes is indicated at the bottom of each panel and the values of the applied potentials (in Volts) are given for three different times in the harmonic waveform cycle. 
The absolute field strength is given for two contour lines in panel (c). 
Panel (d) shows the mechanical potential experienced by the metastable CO molecules in the case of a constant velocity (guiding).  
Panel (e) shows the mechanical potential when a deceleration of 0.78 $\mu$m/$\mu$s$^2$ is applied.
}
\label{principle}
\end{figure}

The operation principle of the Stark decelerator on a chip relies on the superposition of electric fields created by the electrodes on the chip. 
When two dipolar fields with different length scales and opposite directions are superimposed, a minimum of the electric field strength is created. 
This minimum is located at the point where the long range dipole that dominates far from the surface is canceled by the short range dipole that dominates close to the surface. 
Such a minimum of the electric field strength presents a trap for polar molecules in a low field seeking quantum state. 
With the electrode design used in the present study, an array of electric field minima is created above the surface, as shown in Figure \ref{principle}.

The electrode design consists of an extended array of equidistant parallel electrodes with a length of 4 mm arranged on a flat support. 
The electrodes have a width of 10 $\mu$m with a center-to-center separation of neighboring electrodes of 40 $\mu$m. 
This structure is periodically extended over about 50 mm. 
Each electrode is electrically connected to the electrodes that are (multiples of) six positions further, i.e., the electric field repeats itself every 240 $\mu$m. 
Figures \ref{principle}(a)-(c) show two-dimensional plots of the calculated electric field strength for three different sets of potentials. 
The electrodes are indicated at the bottom of each panel by the bold horizontal bars, and the voltages applied to the electrodes are given above them. 
The contours of equal electric field strength are separated by 0.4 kV/cm: the absolute values for two contours are given in panel (c). 
Panel (a) shows a situation where the $z$-position of the minima is exactly between two electrodes. 
The distance of the minimum above the electrodes is primarily determined by the geometric arrangement of the electrodes. 
Further above the substrate, the contour lines run ever more parallel to the surface, and the strength of the electric field decays exponentially with the $y$-position. 
In the region far from the surface, therefore, the electrode array yields a flat, repulsive mechanical potential for polar molecules in low field seeking states. 
Using a slightly different array of electrodes, but using the same underlying principle, a microstructured mirror for polar molecules in low-field seeking quantum states has previously been demonstrated \cite{schulz04}.

Using different sets of potentials, it is also possible to position the minima either directly above an electrode (panel (c)) or in an intermediate $z$-position, three-quarters of the distance from one electrode to the next (panel (b)). 
By applying the appropriate potentials, the minima can be positioned at any $z$-position, while their $y$-position remains constant. 
The periodic arrangement of the electrodes over a length of 50 mm allows for a continuous movement of the electric field minima over this macroscopic distance at a constant height of about 25 $\mu$m. 
As shown in Figure \ref{principle}(a)-(c), six independently variable potentials are required to provide an electric field minimum for every three electrodes. 
It can be shown that it is also possible to obtain movable minima using four independently variable potentials, but in this case there is only an electric field minimum for every four electrodes. 
The temporal variation that is required for each of the six potentials to keep the electric field minima at a constant height ($\Delta y$ $\pm$ 0.05 $\mu$m) while moving at a constant velocity in the $z$-direction turns out to be harmonic. 
Three of the potentials can always be positive, the other three always negative, and within each polarity set the potentials need to be phase-shifted by $120^\circ$. 
Time variation of the potentials with a fixed frequency $\omega$/(2$\pi$), which is in the MHz regime for our experiments, then results in a movement of the minima with a constant velocity $v_z$ given by $v_z = 120\mu\textrm{m} \cdot \omega/(2\pi)$.  

In each of the panels (a)-(c) of Figure \ref{principle}, the times ($t_a$)-($t_c$) at which the contour plots are shown are chosen such that $\omega(t_b - t_a) = \omega(t_c - t_b) = \frac{\pi}{6}$. 
Let the electrodes in panels (a)-(c) be numbered $n = 1$ to $n = 6$ from left to right. 
The voltage at the $n^{th}$ electrode for odd $n$ in panel (a) is given by $-V_0 (1 + \cos(\omega t_a + \phi_n))$, with $V_0 = 60$ V and $(\omega t_a + \phi_n)$ equal to $120^\circ$, $240^\circ$, and $0^\circ$ for $n =$ 1, 3, and 5, respectively. 
For the even numbered electrodes, $n =$ 2, 4, 6, the voltages are positive with $V_0 (1 + \cos(\omega t_a + \phi_n))$ and $(\omega t_a + \phi_n)$ equal to  $0^\circ$, $120^\circ$, and $240^\circ$, respectively. 
In this way, the potential on any given electrode is always equal in magnitude but opposite in sign to the potential on the electrodes that are three positions further.  

\begin{figure}
\center{\includegraphics[scale=0.8]{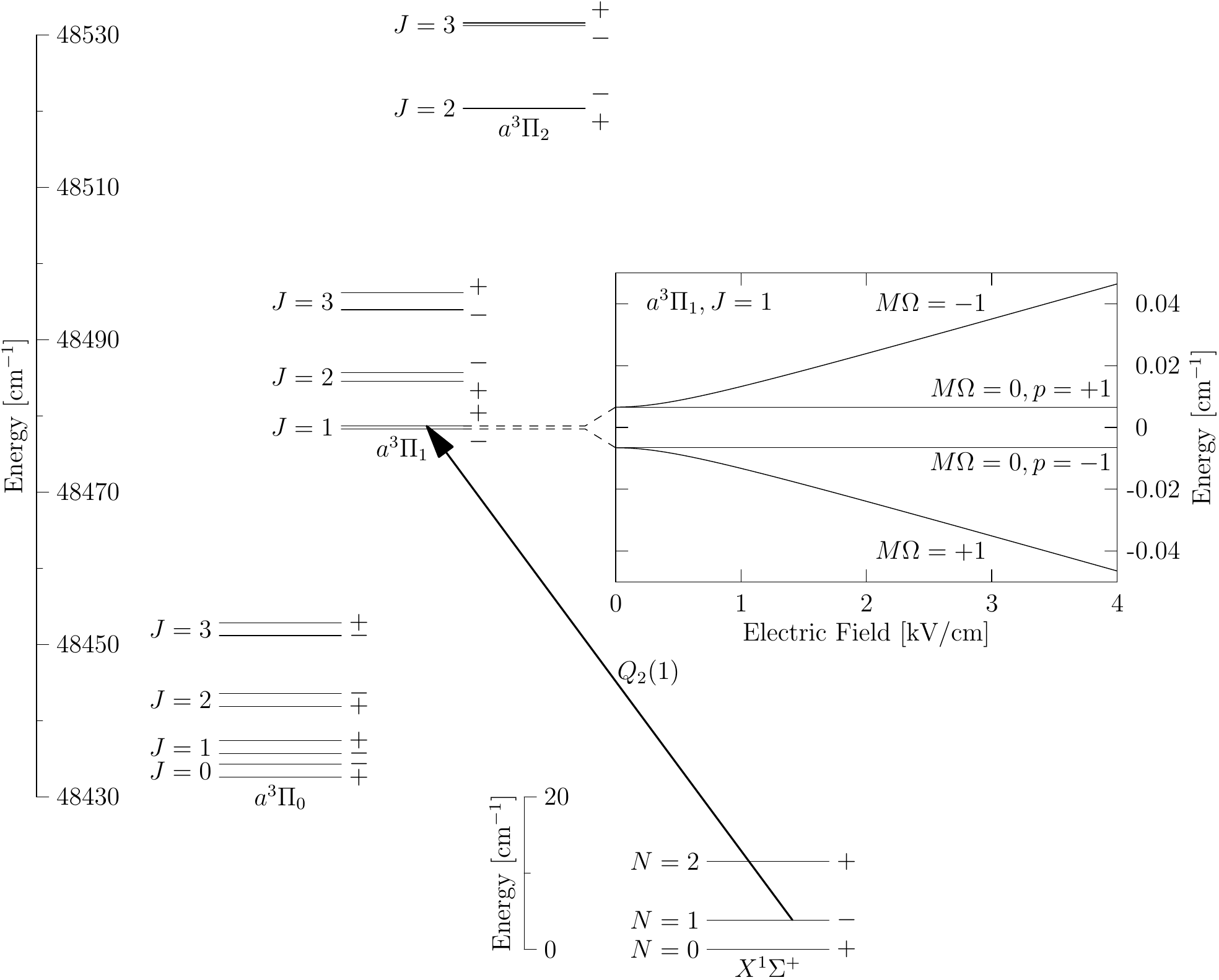}}
\caption{
Rotational energy level structure of the $X ^1\Sigma^+, v"=0$ ground-state and the $a ^3\Pi, v'=0$ metastable state of CO. 
Molecules are prepared in the upper $\Lambda$-doublet component of the $a ^3\Pi_1, v'=0, J' = 1$ level by direct laser excitation on the $Q_2(1)$ transition around 206 nm. 
This spin-forbidden transition becomes weakly allowed via spin-orbit coupling of the $a ^3\Pi_1$ states with higher-lying $A^1\Pi$ states. 
In the inset, the energy of the $a^3\Pi_1, v'=0, J'=1$ level is shown in electric fields up to 4 kV/cm.
}
\label{colevels}
\end{figure}

The Stark deceleration experiments on a chip are performed with CO molecules, laser prepared in the upper $\Lambda$-doublet component of the $a ^3\Pi_1 (v'= 0, J'= 1)$ level. 
This system has also been used in the first demonstration of a Stark decelerator \cite{bethlem99} and in the experiments in which guiding of molecules on a chip was demonstrated \cite{meek08}. 
The relevant energy levels of CO, and the excitation scheme used, are shown in Figure \ref{colevels}. 
This Figure also shows the energy of both $\Lambda$-doublet components of the $a ^3\Pi_1 (v'= 0, J'= 1)$ level in electric fields up to 4 kV/cm. 
Because of the $\Lambda$-doubling, the Stark shift is quadratic for small values of the electric field. 
This leads to slight differences between the shape of the electric field minima and the shape of the mechanical potential experienced by the molecules, especially in the region near zero electric field. 
The potential energy $W(E)$ of the metastable CO molecules in the selected quantum state is given by 
\begin{equation}
W(E) = \sqrt{\left(\frac{\mu E}{2}\right)^2 + \left(\frac{\Lambda}{2}\right)^2} -  \left(\frac{\Lambda}{2}\right)
\end{equation}
where $\mu$ is the body fixed dipole moment (1.37 Debye), $E$ is the magnitude of the electric field and $\Lambda$ is the magnitude of the $\Lambda$-doublet splitting (394 MHz).  
The absolute electric field strength near a minimum can be expressed as
\begin{equation}
E(y,z) = \alpha \sqrt{(y - y_0)^2 + (z - z_0)^2}
\end{equation}
where $y_0$ and $z_0$ are the instantaneous coordinates of the minimum.  
In our choice of coordinate system, where the surface defines $y = 0$, the minima are constantly at $y_0 = 24.7 \mu$m, whereas $z_0$ changes as a function of time. 
The parameter $\alpha$, which describes the slope of the electric field strength near a minimum, is given by 
\begin{equation}
\alpha = 6.0\cdot10^{-4} \mu\textrm{m}^{-2} \cdot V_0 = 3.6 \cdot 10^{-2} \frac{\textrm{V}}{\mu \textrm{m}^2}
\end{equation}
which is more than an order of magnitude larger than the corresponding values for the macroscopic electrostatic quadrupole traps that have been used thus far. 
The resulting potential energy $W(y,z)$ in the region around the center of the well is then given by
\begin{equation}
W(y,z) = \sqrt{\left(\frac{\mu \alpha}{2}\right)^2 ((y - y_0)^2 + (z - z_0)^2) + \left(\frac{\Lambda}{2}\right)^2} 
-  \left(\frac{\Lambda}{2}\right).
\label{walpha}
\end{equation}

Figure \ref{principle}(d) shows the equipotential lines of the well experienced by the metastable CO molecules at 4 mK intervals.  
Along the $z$-axis ($z_0 = 0$ in this case), the potential increases almost linearly with distance from the center, while the potential is more asymmetric perpendicular to the surface. 
At a height of about 50 $\mu$m above the surface there is a saddle point, which limits the effective well depth to about 36 mK. 
The potential only shows quadratic behavior very close to the minimum of the well.  

The described setup can be used to decelerate the molecules captured in the potential well by continuously reducing the frequency of the applied waveforms. 
This reduces the velocity of the potential wells along the surface from their initial velocity, which is matched to the velocity of the incoming molecules, to a certain final velocity while ideally maintaining all trapped molecules within the confines of an individual well. 
During a linear decrease of the frequency with time, the equations of motion are best described in an accelerated reference frame, moving along with the accelerating minima. 
This results in a constant pseudo-force $m \cdot a$ in the positive $z$-direction, opposite to the direction of the acceleration, where $m$ is the mass of the CO molecule and $a$ is the magnitude of the acceleration. 
This pseudo-force can be included in the potential energy $W(y,z)$ by adding a term $-m \cdot a \cdot z$ to the expression given in Equation (\ref{walpha}). 
Choosing a deceleration with a magnitude of $a$ = 0.78 $\mu$m/($\mu$s)$^2$ and using an (arbitrary) offset of the potential such that the minimum near $z=0$ has the potential energy $W(y_0,z_0)=0$ results in the equipotential lines shown in Figure \ref{principle}(e). 
It is observed in this contour plot that the position of the minimum is slightly shifted to higher $z$-values and, more significantly, that the line connecting the minimum with the saddle point is strongly rotated forward relative to the constant velocity case. 
This brings the saddle point to substantially lower $y$-values and reduces the depth of the effective potential well for the metastable CO molecules to about 16 mK.

Before concluding this section, two points need to be mentioned. 
First, since a major part of the potential well is linear in both the $y$- and the $z$-direction, the $y$- and $z$-motions of individual trapped molecules are strongly coupled. 
As a result, the equations of motion cannot be solved nor approximated analytically, and numerical calculations need to be performed to get a more detailed understanding of the motion of the trapped molecules. 
Secondly, the operation principle of the Stark decelerator as presented here is fundamentally different from that of the conventional Stark \cite{bethlem99} or Zeeman \cite{vanhaecke07} decelerators. 
In the latter decelerators, switching between two static configurations generates an {\it effective} potential well that travels at a gradually decreasing velocity. 
This results in the occurrence of overtone modes, and in specific couplings of the transverse with the longitudinal motion that can limit the acceptance in these decelerators \cite{meerakker06}. 
In the decelerator concept presented here we have a genuine traveling potential well, which will prevent some of the stability problems inherent in other decelerators.

\section{Experimental setup}

\begin{figure}
\center{\includegraphics[scale=0.8]{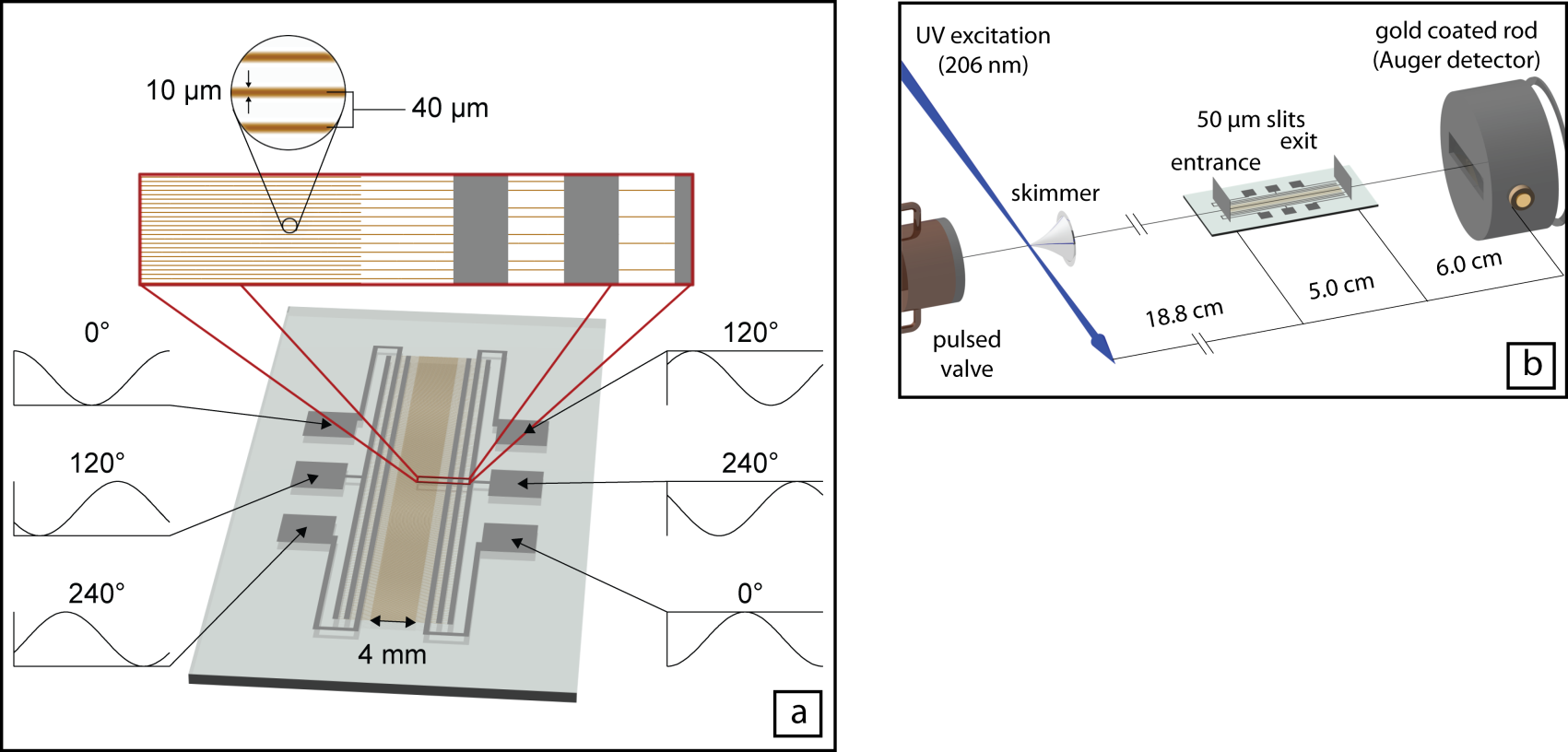}}
\caption{
(a): Picture of the chip decelerator in three stages of magnification. 
The waveforms that are applied to the pads are schematically shown. 
(b): Scheme of the experimental setup with the relevant distances along the beamline indicated.
}
\label{expset}
\end{figure}

The physical realization of the decelerator on a chip is shown in Figure \ref{expset}(a). 
An array of 1254 10-$\mu$m wide and approximately 100 nm high gold electrodes are deposited onto a glass substrate with a 40 $\mu$m center-to-center spacing, forming a decelerator structure that is 50.16 mm long (micro resist technology GmbH). 
All electrodes extend over a central 4 mm region, and outside this region, the electrodes extend alternatingly to the left or the right and terminate at three different lengths. 
The electrodes are covered with a 200 nm thin protecting and insulating layer (SU-8), except near their tips. 
This allows the nickel connecting wires on either side to connect to every third electrode on that side, while still being insulated from the other two sets of electrodes. 
The nickel electrodes in turn connect to square pads, via which the six different waveforms can then be applied to the microstructured electrodes. 
The SU-8 layer has an electrical permittivity of $\epsilon$=4, and its presence has been taken into account in the calculated electric field distributions shown in Figure \ref{principle}. 
It is seen to have only a slight effect on the electric fields in vacuum, but it effectively prevents free electrons from reaching the electrodes and 
producing discharges.  

A schematic view of the beam machine with the chip decelerator is shown in Figure \ref{expset}(b). 
The chip with the decelerator is mounted on a manipulator that allows independent {\it in situ} adjustment of the height of the front and the back of the decelerator with a precision of better than 10 $\mu$m. 
These two linear adjustments also enable the structure to be tilted with a precision of better than $0.01^\circ$. 
Additional adjustments prior to mounting the system in vacuum allow for rotation and translation of the structure in the plane of the surface. 
By removing the pulsed valve and shining a He-Ne laser along the path of the molecular beam, it is possible to accurately align the microstructured electrodes of the decelerator perpendicular to the molecular beam. 

In the experiment, a pulsed beam (10 Hz) of ground-state CO ($X^1\Sigma^+$, v"=0) molecules is produced by expanding a 1.2 bar mixture of 20\% CO in argon through a solenoid valve (General Valve; series 99) in vacuum. 
The valve body is cooled to 102 Kelvin to reduce the mean velocity of the CO molecules to 360 m/s. 
Just before passing through the skimmer, CO molecules are prepared in the upper $\Lambda$-doublet component of the $a^3\Pi_1 (v'=0, J'=1)$ level by direct laser excitation from the $N"=1$ rotational level in the ground-state. 
Using a 5 ns duration laser pulse at 206 nm with 1 mJ of energy and with a Fourier-transform limited bandwidth, this spin-forbidden transition can actually be saturated even when the laser beam is only weakly focused (1 mm diameter) onto the molecular beam. 
The excitation is performed in a region without an applied external field and, as $^{12}C^{16}O$ has no hyperfine splitting, the $a^3\Pi_1 (v'=0, J'=1)$ level is three-fold degenerate.
As is shown in the inset in Figure \ref{colevels}, this three-fold degenerate level will split into two sublevels in an electric field: a nondegenerate $M = 0$ level which has only a weak, negative quadratic Stark shift and a degenerate pair of $M \Omega = -1$ levels that has a linear positive Stark shift already at relatively low electric fields. 
Only molecules in the low field seeking degenerate pair of $M \Omega = -1$ levels can be trapped or deflected with the electric fields above the chip, whereas the trajectories of molecules in the $M = 0$ level will be almost unaffected by the fields. 
The $a ^3\Pi_1 (v'= 0, J'=1)$ level has a radiative lifetime of 2.63 ms \cite{gilijamse07}.
With a typical time-of-flight of the molecules through the compact beam machine of about 1 ms, this implies that over two-thirds of the molecules will reach the detector while still in the metastable state. 
At the same time, this finite lifetime permits the recording of some laser induced phosphorescence signal just after the molecules have passed through the 1.0 mm diameter skimmer into the second chamber; i.e., it allows for monitoring the beam intensity and for optimizing the excitation laser frequency. 

About 19 cm downstream from the laser excitation point, the molecules pass through an approximately 50 $\mu$m high entrance slit and then travel closely above the microstructured electrode array over its full 5 cm length. 
When the molecules have just arrived above the electrode array, the external potentials are switched on. 
The switched, phase-stable waveforms needed for this experiment are generated using an arbitrary function generator (Aquitek DA8150). 
This card produces up to eight independent waveforms that are each generated by reading a series of 12-bit values at a sample rate of 150 MHz and converting these to an analog signal between -1 V and +1 V. 
These analog signals are then amplified by home-built push-pull tube amplifiers that have a bandwidth from about 3 MHz down to DC. 
The exact time that the waveforms are switched on is 530 $\mu$s after the excitation. 
The position of laser excitation is 188 mm from the first electrode of the decelerator, so molecules traveling at the mean velocity of 360 m/s have then already passed over 3 mm of decelerator electrodes, whereas molecules with a velocity of 355 m/s have just reached the first electrodes. 
A small fraction of the molecules will find themselves into the electric field minima, while the majority will be deflected away from the surface by the strong electric field gradient that is present everywhere else. 
When the electric field minima move at the right initial velocity, the metastable CO molecules can stay in the minima for the entire time that the waveforms are applied. 
When the molecules are close to the end of the electrode array, the external potentials are switched off again. 
The exact time during which the waveforms are kept on depends on the amount of deceleration that is imposed, but it is always chosen such that it corresponds to a total travel distance of the guided or decelerated molecules above the chip of 46 mm. 
The molecules that have been stably transported over this 46 mm distance from one end of the chip to the other can pass through a 50 $\mu$m high exit slit and then fly freely to a gold surface that is positioned 6 cm further downstream. 
As the metastable CO molecules have 6 eV of internal energy, they can emit Auger electrons when impacting the gold surface, and these electrons are subsequently detected using a multi-channel plate detector. 
The efficiency with which a metastable CO molecule is detected is estimated to be on the order of 1\% in this scheme.  

\section{Experimental results}

\begin{figure}
\center{\includegraphics[scale=0.8]{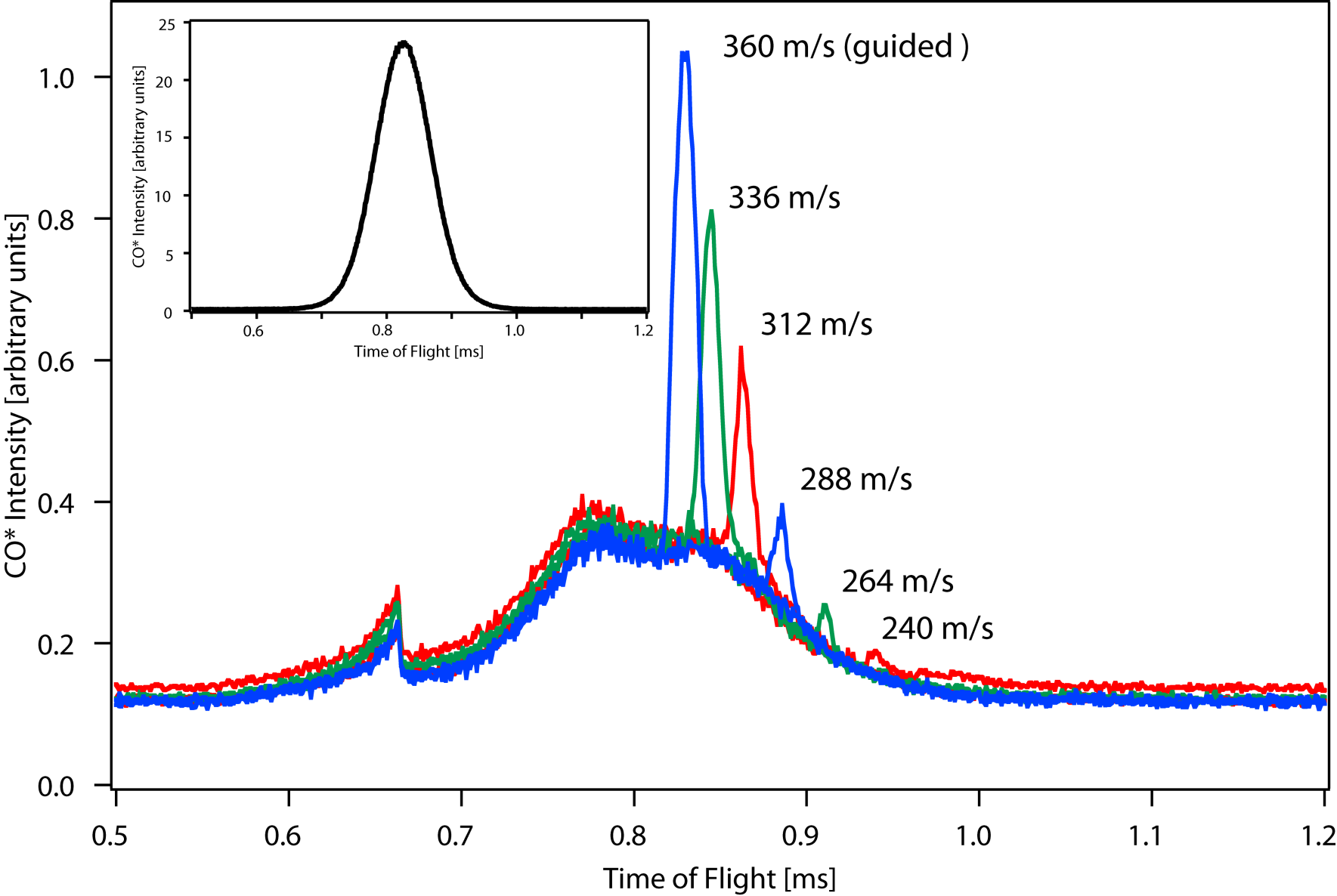}}
\caption{
Measured arrival time distributions of metastable CO molecules for six different acceleration values.
In all measurements, molecules with an initial velocity of 360 m/s are brought to the final velocity that is indicated next to the main peaks in the arrival time distribution. 
Inset: arrival time distribution when no voltages are applied to the microstructured electrodes. 
All curves are shown on the same vertical scale.
}
\label{expdata}
\end{figure}

The experimentally measured arrival time distributions of the metastable CO molecules are shown in the main panel of Figure \ref{expdata} for six different values of the applied acceleration. 
In the inset, the measured arrival time distribution of metastable CO molecules is shown when no potentials are applied to the electrodes. 
From this measurement it is seen that the velocity distribution of the CO molecules in the beam is centered at 360 m/s and that it has a relatively large full width at half maximum of about 45 m/s. 
When the waveforms are switched on, they have an initial frequency $\omega/(2 \pi)$ of 3.0 MHz, implying that the potential wells initially move with a velocity of 360 m/s. 
It was already mentioned in the previous section that the waveforms are switched on at the time when molecules moving with an initial velocity of 360 m/s have just arrived on the decelerator. 
CO molecules that are not initially in a potential well will be deflected away from the surface, and even molecules that are initially in the potential wells but that have a large velocity relative to the wells will soon find their way out and will then be pushed away from the surface. 
In the guiding case, where the waveforms remain at 3.0 MHz for the entire duration that the molecules are on the chip, molecules in the wells are carried to the other end of the chip without their velocity changing significantly. 
The resulting time of flight spectrum shows a broad background on top of which there is an intense, narrow peak centered at 0.83 ms, i.e., at the time it takes to travel with a constant velocity of 360 m/s over the full 29.8 cm distance from the excitation zone to the detector. 
The broad background results from metastable CO molecules that are hardly influenced by the electric fields, most likely from molecules in the $M$=0 level. 
The discontinuity that is seen in the background at an arrival time of around 0.66 ms results from the switching on of the waveforms: fast CO molecules that exit the decelerator just before the fields are switched on reach the detector unperturbed, whereas the slower molecules that are still above the surface when the fields are switched on are deflected away from the surface and do not reach the detector anymore. 
Although the intensity scale in Figure \ref{expdata} is in arbitrary units, the same vertical scale is used for all curves. 
It is thus seen that only about 4\% of the molecules with an initial velocity around 360 m/s that pass in free flight through both of the 50 $\mu$m high slits is actually captured in the electric field minima in the case of guiding. 
Considering the size of the minima in this case (see Figure \ref{principle}(d)), this is about what would be expected purely based on geometrical arguments. 

The other five arrival time distributions shown in the main panel of Figure \ref{expdata} are recorded when the frequency of the waveforms is reduced linearly in time from the initial value of 3.0 MHz, to a final value of 2.8 MHz (336 m/s), 2.6 MHz (312 m/s), 2.4 MHz (288 m/s), 2.2 MHz (264 m/s) and 2.0 MHz (240 m/s). 
The last of these measurements corresponds to a deceleration with a magnitude of $a$ = 0.78 $\mu$m/($\mu$s)$^2$, for which the shape of the mechanical potential well was shown already in Figure \ref{principle}(e). 
It is seen that the narrow peak in the arrival time distribution that results from the metastable CO molecules that are captured in the potential wells comes at ever later times, in perfect agreement with the expected arrival time, given the geometry of our experimental setup ({\it vide infra}). 
The broad background, on the other hand, is more or less constant and hardly depends on the frequency of the applied potentials.

It is seen from the measurements shown in the main panel of Figure \ref{expdata} that the number of molecules that is trapped in the potential wells decreases for increasing accelerations. 
This is expected because the volume of the potential wells gets smaller and, in addition, the wells become less deep with increasing accelerations. 
Trajectory calculations are required, however, to be able to conclude whether or not this reduction in the phase-space acceptance of the potential wells with increasing acceleration can quantitatively account for the experimentally observed reduction in intensity with increasing acceleration.

\section{Trajectory calculations}

\begin{figure}
\center{\includegraphics[scale=0.8]{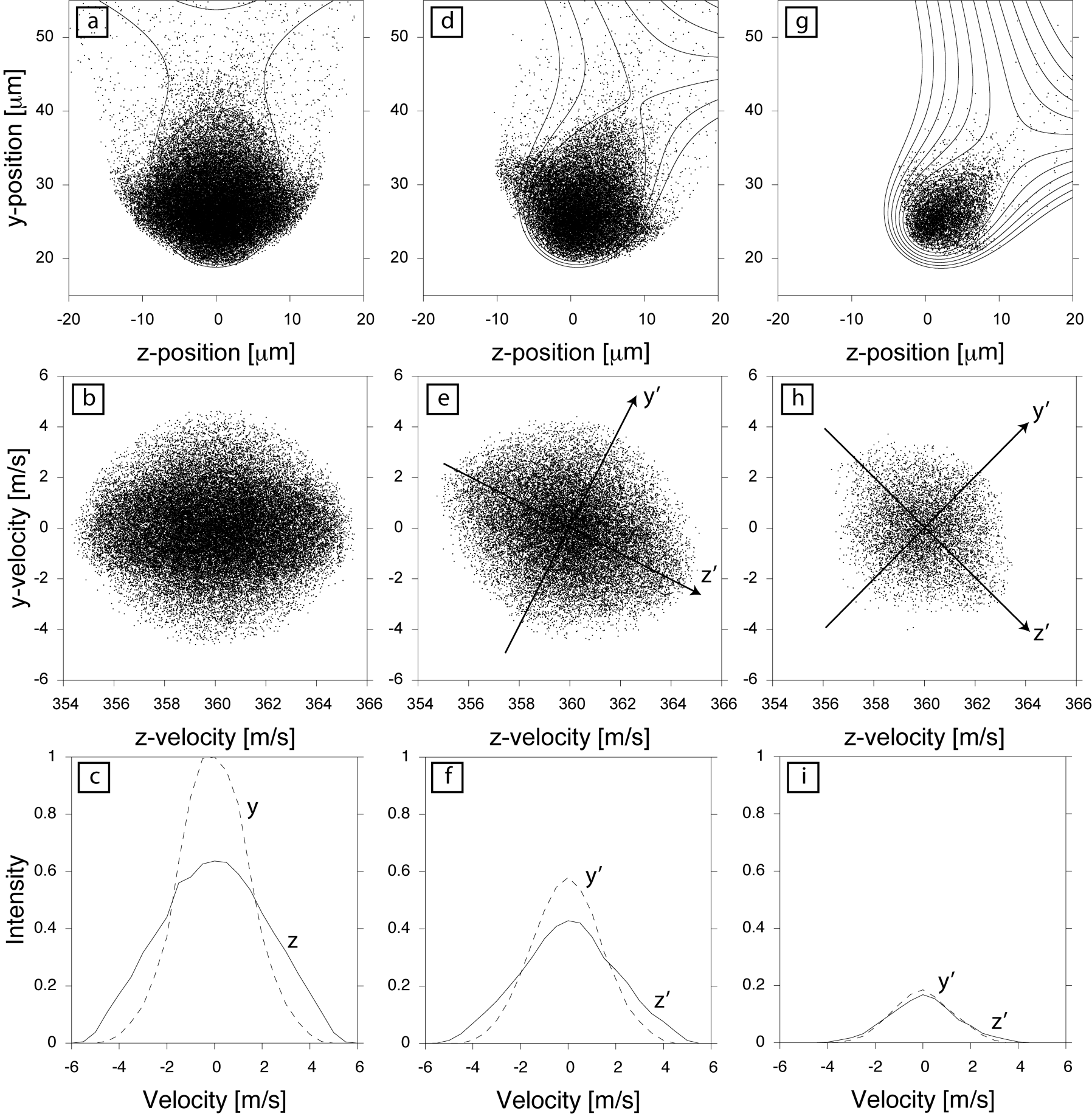}}
\caption{
Calculated initial positions and velocities of metastable CO molecules in the low field seeking component of the $a^3\Pi_1, v'=0, J'=1$ level that reach the detector when guided at a constant velocity of 360 m/s ((a)-(c)), when decelerated from 360 m/s to 312 m/s ((d)-(f)), and when decelerated from 360 m/s to 240 m/s ((g)-(i)). 
Each simulation uses a homogeneous distribution of positions and velocities over a hyperrectangle of 50 $\mu$m by 50 $\mu$m by 20 m/s by 20 m/s, centered at $y$ = 30 $\mu$m, $z$ = 0 $\mu$m, $v_y$ = 0 m/s, and $v_z$ = 360 m/s.}
\label{minpts}
\end{figure}

To gain a better insight into the motion of the molecules in the decelerator, two-dimensional numerical trajectory simulations are carried out. For this, the electric potentials in the area near the surface are calculated using a commercially available finite-element program (COMSOL) assuming a strictly periodic array of electrodes. In these potential calculations, both the dielectric constants of the SU-8 and that of the glass underneath have been accounted for. 
It is straightforward to simulate the fields produced by an arbitrary set of electrode potentials by carrying out the calculation with 1 V applied to one electrode while the other electrodes are grounded. 
The calculated potential can be shifted in the $z$-direction by one or more electrode spacings and the total potential can be calculated by forming a linear combination of these shifted potentials.

A two-dimensional electric potential that is periodic in $z$ and that vanishes as $y \rightarrow \infty$ can be written as
\begin{equation}
V(y,z) = \sum_{n = 1}^{\infty} V_n \cos \left(\frac{2 \pi n z}{l} + \delta_n\right) e^{-\frac{2 \pi n y}{l}}
\label{Vyz}
\end{equation}
where $l$ is the periodicity of the potential ($l$ = 240 $\mu$m in our case).
The coefficients $V_n$ and the phases $\delta_n$ can then be determined by expanding the numerically calculated potential at a constant value of $y$ in the Fourier series. By choosing a $y$-value close to the surface, these parameters are obtained with the highest accuracy, since the numerical calculation of the electric potential is most accurate near the electrodes. The Fourier series includes contributions that result from numerical noise that is present in the finite-element calculations, but because these contributions are predominantly in the high-$n$ terms, the effect of the numerical noise becomes negligible at $y$ = 25 $\mu$m, where the molecules are. The analytical fitting thus has the added benefit of smoothing the short-range roughness of the numerically calculated potential.

Using this analytic form for the electric potential, the mechanical potential can be calculated as a function of position and time, and with that, numerical trajectory simulations have been performed.  
One set of simulations uses $10^7$ initial positions evenly distributed over phase space in a hyperrectangle 50 $\mu$m by 50 $\mu$m by 20 m/s by 20 m/s, centered at $y$ = 30 $\mu$m, $z$ = 0 $\mu$m, $v_y$ = 0 m/s, and $v_z$ = 360 m/s. 
Figure \ref{minpts} shows the subset of initial positions and velocities of the molecules that reach the detector after being guided or decelerated over the entire length of the chip, for the final velocities 360 m/s ((a)-(c)), 312 m/s ((d)-(f)), and 240 m/s ((g)-(i)).  
In \ref{minpts}(a), \ref{minpts}(d), and \ref{minpts}(g), the positions of most of the molecules that are accepted are near the minimum of the effective potential with potential energies below the barrier. 
Some molecules that are outside this region have metastable trajectories, and would not reach the detector if the decelerator were longer.  The points in the lobes perpendicular to the barrier, however, generally correspond to stable trajectories; although the total mechanical energy of these molecules is sufficient to escape, their trajectories are such that they never find the exit. 
The velocity distributions, shown in \ref{minpts}(b), \ref{minpts}(e), and \ref{minpts}(h), have a wider range of acceptance perpendicular to the barrier (along the $z'$-direction) than in the direction towards the barrier (along the $y'$-direction);
the velocity distributions along these axes are shown in \ref{minpts}(c), \ref{minpts}(f), and \ref{minpts}(i). Due to the reduced depth of the effective potential well, the velocity acceptance decreases for increasing deceleration.

\begin{figure}
\center{\includegraphics[scale=0.8]{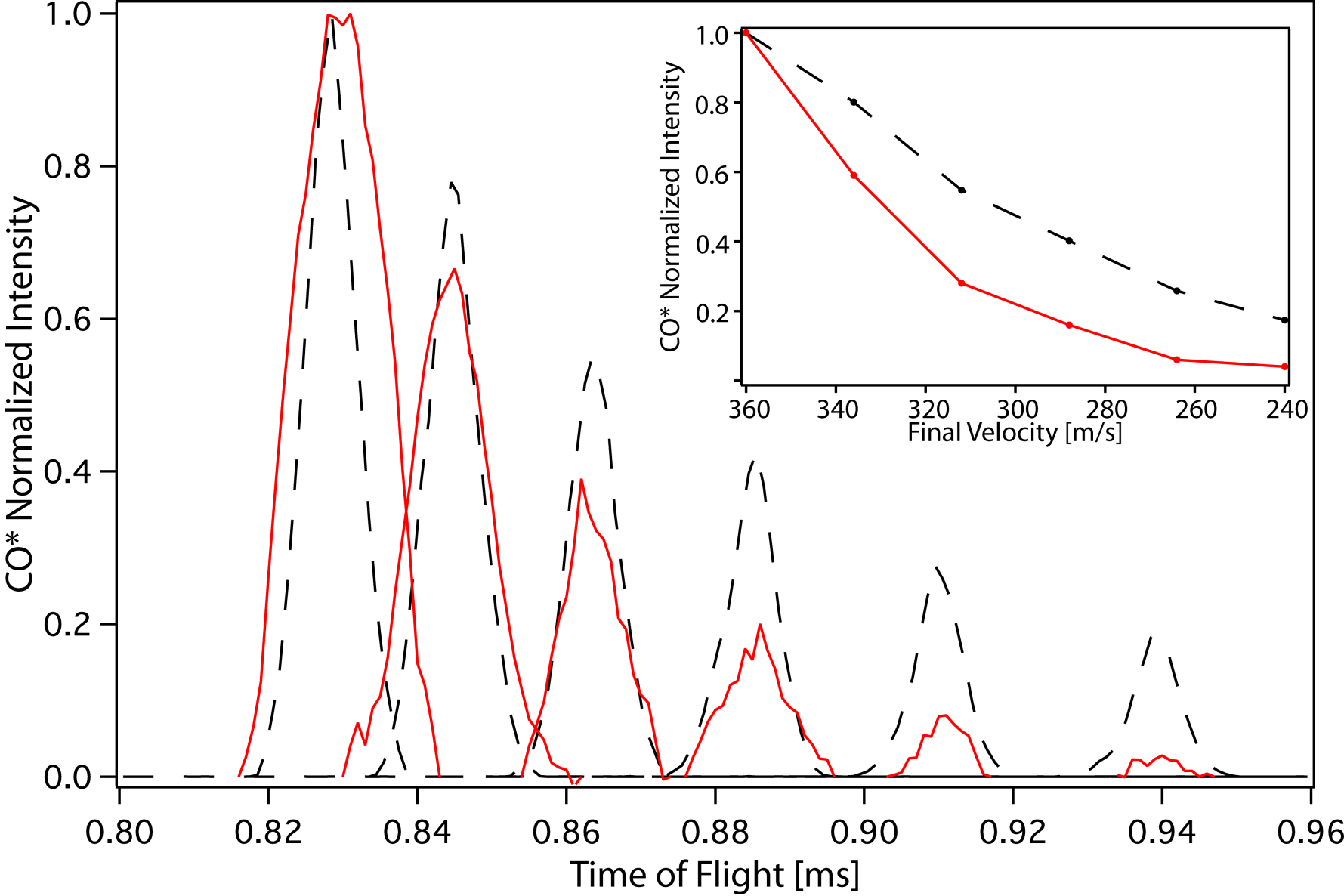}}
\caption{
Comparison between the measured (solid red) and the calculated (dashed black) arrival time peaks for the parameter sets used for the measurements shown in Figure \ref{expdata}. 
The measured peaks are obtained by subtracting the background. 
Both sets are normalized to a peak intensity of one for the guided peak. 
The inset shows the integrated intensity, normalized to one for the integrated guided intensities, as a function of the different final velocities.
}
\label{exptheo}
\end{figure}

Another set of trajectory calculations were done to simulate the arrival time distributions. For this, a molecule is first generated randomly on a circular aperture, corresponding to the position of the skimmer.  
It is given a velocity $v_z$ from a Gaussian distribution centered at 360 m/s with a full width at half maximum of 45 m/s, and a velocity $v_y$ from a flat distribution, chosen such that the molecule passes through the 50 $\mu$m high entrance slit at the front of the decelerator.  
The molecule then flies freely until it arrives above the decelerator and the fields are switched on.  
While the fields are on, the trajectory is calculated as described before.  
After the fields are switched off or when the molecule leaves the decelerator, it propagates freely to the position of the exit slit, and if its $y$-position is then less than 50 $\mu$m from the surface, it continues propagating freely to the detector, and its total flight time is recorded.  

For each of the six deceleration sequences measured, $10^6$ trajectories were simulated.  
The resulting arrival time distributions are shown in Figure \ref{exptheo} as dashed black curves, 
and their positions are seen to agree very well with those of the corresponding experimental peaks, 
shown as solid red curves. In the calculations it is assumed that all trajectories start at the same 
$z$-position, whereas the experiment has a range of initial $z$-positions due to the finite width of 
the laser beam, which is about 1 mm.  As a result, the calculated peaks are a bit narrower than those 
in the experiment.  The total intensity in the experiment, however, is seen to decrease much more 
rapidly with increasing deceleration than the calculations predict.  

There are a number of possible factors that could contribute to this discrepancy. Since the calculations are done in two dimensions, they neglect the molecules that escape to the side (in the $x$-direction) of the decelerator.  
For a given initial velocity, the number of molecules reaching the detector will be inversely proportional to their arrival time due to this effect. Losses also occur due to the finite lifetime of the metastable state, and these losses will be higher for lower final velocities. The combined effect of these two loss processes only yields an additional 15\% loss of the most decelerated peak relative to the guided peak. This has not been taken into account in the calculations, but is by
far not sufficient to explain the observed discrepancy.
Imperfections in the geometry of the electrodes and in the waveforms applied to them could also cause the minima to move erratically, effectively making the potential wells shallower. It might well be, that collisions of the carrier gas atoms with the metastable CO molecules eject trapped molecules from the potential wells. 
This could result in greater losses for the more decelerated molecules, since these are in a shallower trap and have a larger velocity relative to the carrier gas.
Finally, non-adiabatic transitions to the $M=0$ level that might occur when the molecules pass near the zero-field region of the potential wells cannot be completely ruled out. This effect is also expected to result in greater losses when the size of the potential well decreases.

\section{Conclusions and outlook}

In this paper, we have experimentally demonstrated a Stark decelerator on a chip. 
This decelerator can be loaded directly from a molecular beam and can then be used to either guide the molecules at a constant velocity along the chip, or it can be used to decelerate them while they are above the chip. 
Using metastable CO molecules, deceleration from an initial velocity of 360 m/s to a final velocity of 240 m/s has been demonstrated, thereby removing more than 50\% of the initial kinetic energy.

In the work presented here, argon has been used as a seed gas, thereby producing a beam of CO molecules with a mean velocity of 360 m/s. 
If xenon is used instead, the initial velocity is reduced to 300 m/s, which, with the already demonstrated maximum deceleration, would enable us to bring the CO molecules to a final velocity of 130 m/s on the chip. 
With a slightly higher deceleration, made possible by somewhat higher applied potentials, the molecules could be brought to standstill on the present chip.  

Two-dimensional numerical trajectory calculations, simulating the full geometric arrangement, show excellent agreement with the experimental data, although the observed reduction of the number of decelerated molecules with increasing deceleration is higher than expected from the calculations. 
The additional loss that is experimentally observed might result from elastic collisions between the metastable CO molecules and the rare gas atoms in the beam, effectively kicking the molecules out of the rather shallow traps. 
To mitigate this potential problem, we will install a second identical chip between the skimmer and the decelerator. 
The electrodes on this additional chip will be configured such that it acts as an electrostatic mirror for polar molecules. 
With this mirror, only metastable CO molecules in a the low field seeking state will be focused through the entrance slit onto the decelerator, whereas the rare gas atoms as well as CO molecules in the $M=0$ level will be diffusely scattered. 

\section*{Acknowledgments}

We acknowledge useful discussions with Andreas Osterwalder and Hendrick L. Bethlem. The design of the electronics by G. Heyne, V. Platschkowski, and T. Vetter has been crucial for this work.
This work has been funded by the European Community's Seventh Framework Program FP7/2007-2013 under grant agreement 216 774.

\end{document}